\pdfoutput=1	

\documentclass[final,5p,times,twocolumn]{elsarticle}

\usepackage{amsmath}
\usepackage{bm}
\usepackage{hyperref}

\journal{Optics Communications}

\begin{document}

\begin{frontmatter}

\title{Law of refraction for generalised confocal lenslet arrays}

\author{Stephen Oxburgh}
\ead{s.oxburgh.1@research.gla.ac.uk}
\author{Chris D.\ White}
\author{Georgios Antoniou}
\author{Johannes Courtial}
\ead{Johannes.Courtial@glasgow.ac.uk}
\address{SUPA, School of Physics \& Astronomy, University of Glasgow, Glasgow G12~8QQ, United~Kingdom}

\begin{abstract}
We derive the law of generalised refraction for generalised confocal lenslet arrays, which are arrays of misaligned telescopes.
We have implemented this law of refraction in TIM, a custom open-source ray tracer.
\end{abstract}


\end{frontmatter}

\newcommand{\rmi}{\mathrm{i}}
\newcommand{\rmd}{\mathrm{d}}
\newcommand{\bi}[1]{\bm{#1}}
\newcommand{\p}{^\prime}


\section{Introduction}

\noindent
The recent realisation using metamaterials of an interface that changes the direction of transmitted light rays according to a ``law of generalised refraction'' that is different from Snell's law \cite{Yu-et-al-2011} has significantly raised the profile of generalised refraction.
Our own interest in generalised refraction originates from our work on METATOYs \cite{Hamilton-Courtial-2009}, which are structured sheets that change the direction of transmitted light rays in unusual ways, including reflection off a plane other than the interface \cite{Hamilton-Courtial-2008a} and rotation around arbitrary axes \cite{Hamilton-et-al-2009,Hamilton-et-al-2010}.
METATOYs also introduce an offset into transmitted light rays, and indeed they have to do this:
without such an offset the refraction performed by METATOYs would be --- at least for some incident light fields \cite{Courtial-Tyc-2012} --- forbidden by wave optics.
The idea is that the offset can be made sufficiently small so that it is virtually unnoticeable in certain situations.

We consider here a very versatile METATOY called generalised confocal lenslet arrays (gCLAs), which have seven degrees of freedom \cite{Hamilton-Courtial-2009b}.
So far, the law of refraction (we drop the ``generalised'' from now on) was published only for two-dimensional gCLAs \cite{Hamilton-Courtial-2009b}; the law of refraction for three-dimensional gCLAs was calculated \cite{Hamilton-2010a}, but not found to be of sufficient interest to be published.

We recently found potential applications for gCLAs, and we require the law of refraction in a form suited to our purposes.
Here we describe a parametrisation of gCLAs that is (slightly) different from that used previously, and we derive the generalised law of refraction for gCLAs parametrized in this way.
We have also extended our open-source raytracer TIM \cite{Lambert-et-al-2012,Oxburgh-et-al-2013} by adding the capability to trace rays through gCLAs, and we demonstrate TIM's new capability.

\section{Confocal lenslet arrays and generalised confocal lenslet arrays}

\noindent
Two lenses, one placed behind the other such that the front lens's back focal plane coincides with the back lens's front focal plane, form a telescope.
Two arrays of lenses, one placed behind the other such that their focal planes similarly coincide, form an array of telescopes.
They can be built from standard lenslet (or microlens) arrays, which is why they are called confocal lenslet arrays (CLAs) \cite{Courtial-2008a} (Fig.\ \ref{CLAs-ua-projection-figure}).
Almost irrespective of where a light ray hits \cite{Courtial-2009}, CLAs re-direct light rays according to a generalised law of refraction \cite{Courtial-2008a}
\begin{equation}
\tan \alpha_1 = \eta \tan \alpha_2,
\label{tan-refraction-equation}
\end{equation}
where $\alpha_1$ and $\alpha_2$ are the angles of incidence and refraction and
\begin{equation}
\eta = - \frac{f_2}{f_1},
\end{equation}
where $f_1$ and $f_2$ are the focal lengths of the first and second lenslet arrays.
CLAs also offset transmitted light rays on the scale of an individual telescope's aperture radius, but by miniaturising the telescopes this offset can also be miniaturised (but not too much, as diffraction then dominates).
Viewed from a suitable distance from which the individual telescopes cannot be resolved, CLAs appear to be light-ray-direction changing windows, which makes them examples of METATOYs \cite{Hamilton-Courtial-2009}.
These properties of CLAs have also been demonstrated experimentally~\cite{Courtial-et-al-2010}.

Surfaces that refract according to Eqn (\ref{tan-refraction-equation}) (and which do not offset transmitted light rays) have interesting imaging properties:
\begin{enumerate}
\item A planar surface of this type images all space, ray-optically perfectly, with transverse magnification $+1$ and longitudinal magnification $\eta$ or $\eta^{-1}$, depending on the direction with which light passes through the surface \cite{Courtial-2008a}.
\item The combination of an idealised thin lens and planar surface of this type is the most general planar surface that images all space (and which does not offset light rays) \cite{Courtial-2009b}.
\end{enumerate}
CLAs are arguably the best realisation of surfaces that refract according to Eqn (\ref{tan-refraction-equation}).
Perhaps the reason why such surfaces have so far not been realised without the ray offset is that Eqn (\ref{tan-refraction-equation}) is an example of a generalised law of refraction that is wave-optically forbidden for some incident light fields~\cite{Courtial-Tyc-2012}.

CLAs can be generalised without losing their METATOY character, resulting in generalised CLAs (gCLAs) \cite{Hamilton-Courtial-2009b}.
The generalisation works by modifying the individual telescopes such that they retain their property that incident rays that are parallel remain so after transmission through the telescope.
The modifications are as follows:
\begin{enumerate}
\item The two lenses that form an individual telescope in gCLAs can be moved relative to each other in the direction perpendicular to the lenses' optical axis, such that the optical axes of the two lenses no longer coincide, but remain parallel.
\item Each lens can be replaced by a pair of cylindrical lenses with different focal lengths and whose cylinder axes (the curved surface of a cylindrical lens is a section of a cylinder) are perpendicular to each other.
The other lens of each telescope has to be replaced by a matching pair of cylindrical lenses, such that the separation between cylindrical lenses with the same cylinder-axis orientation is the sum of their focal lengths.
\item Each telescope can be rotated such that its axis is no longer perpendicular to the (tangent) plane of the gCLAs.
\end{enumerate}

As a result of these generalisations, gCLAs have 7 degrees of freedom, which makes them very versatile.
We are currently investigating the use of gCLAs in a number of different contexts, and this requires knowledge of the generalised law of refraction for gCLAs.
We derive such a generalised law of refraction here, using a slight variation of the original parametrisation of gCLAs, which results in a form of the generalised law of refraction that is more suitable to our purposes.
Note that a generalised law of refraction for gCLAs, using the original parametrisation, had been derived previously \cite{Hamilton-2010a}.
We have also programmed the generalised law of refraction into our custom ray tracer TIM \cite{Lambert-et-al-2012,Oxburgh-et-al-2013}, which allows visualisation of the view through different gCLAs.

\section{Alternative parametrisation of gCLAs}

\noindent
In Ref.\ \cite{Hamilton-2010a}, each individual telescope in such gCLAs was described in the following way.
Initially, each telescope was orientated such that the $z$ direction was the optical axis, and the cylindrical lenses were orientated such that the cylinder axis of one pair of cylindrical lenses was in the $x$ direction, that of the other pair was in the $y$ direction.
The cylindrical lenses whose axes are orientated in the $x$ direction, in the idealisation we use here, affect only the $y$ component of the direction of transmitted light rays, and so the pair of cylindrical lenses that orientated in this way is characterised in terms of focal lengths $f_{1, y}$ and $f_{2, y}$.
What matters for the light-ray direction change is only the ratio of these focal lengths, which is described by the parameter $\eta_y = - f_{2, y} / f_{1, y}$.
Similarly, the ratio of focal lengths of the other pair of cylindrical lenses is described by a parameter $\eta_x = - f_{2, x} / f_{1, x}$.
Because the optical axis is in the $z$ direction, the displacement of the optical axes relative to each other is in the $x$ and $y$ directions.
These displacements are characterised by the parameters $\delta_x = d_x / f_{1, x}$ and $\delta_y = d_y / f_{1, y}$, where $d_x$ and $d_y$ are the actual displacements in the $x$ and $y$ directions.
To achieve full generality, the telescopes were then rotated.
This rotation was completely general and described in terms of three rotations, first a rotation by an angle $\varphi$ around the $z$ axis, followed by a rotation by $\theta$ around the $x$ axis, followed by a rotation by $\psi$ around the $z$ axis again, i.e.\ an Euler rotation characterised by the Euler angles $(\varphi, \theta, \psi)$.

Here we describe the same individual telescopes slightly differently.
We define the vectors $\hat{\bm{u}}$, $\hat{\bm{v}}$ and $\hat{\bm{a}}$, which correspond to the directions the vectors $\hat{\bm{x}}$, $\hat{\bm{y}}$ and $\hat{\bm{z}}$ would point in after Euler rotation characterised by the Euler angles $(\varphi, \theta, \psi)$.
The vectors $\hat{\bm{u}}$ and $\hat{\bm{v}}$ therefore point in the directions of the cylinder axes of the cylindrical lenses, and the vector $\hat{\bm{a}}$ points in the direction of the lenses' optical axes.

\section{Generalised law of refraction}

\begin{figure}
\begin{center}
\includegraphics{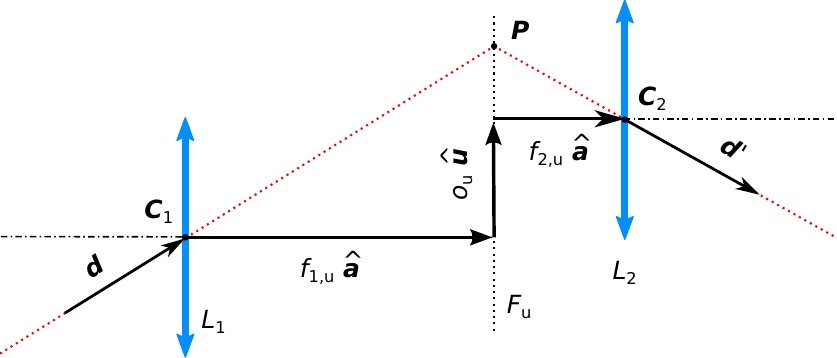}
\end{center}
\caption{\label{CLAs-ua-projection-figure}Geometry of the light-ray-direction change on transmission through one of the generalised telescopes that form generalised confocal lenslet arrays (gCLAs) \cite{Hamilton-Courtial-2009b}, and therefore of the generalised law of refraction for gCLAs.
The drawing shows an orthographic projection into the $(u, a)$ plane, i.e.\ the plane spanned by the common direction of the optical axes of the lenses, $\hat{\bm{a}}$, and the one of the two cylinder-axis directions, $\hat{\bm{u}}$.
$L_1$ and $L_2$ are the telescope lenses in the order in which an incident light ray with direction $\bm{d}$ encounters them.
The direction of the outgoing light ray is given by the vector $\bm{d}^\prime$.
$\bm{C}_1$ and $\bm{C}_2$ are the centres of the lenses; $F_u$ is the plane a distance $f_{1,u}$ behind $L_1$ and a distance $f_{2,u}$ in front of $L_2$; and $o_u$ is the component in the $u$ direction of the offset between the optical axes of $L_1$ and $L_2$.
}
\end{figure}

\noindent
In this derivation of the law of refraction for gCLAs we broadly follow Ref.\ \cite{Hamilton-2010a}.
We study light-ray transmission through one telescope generalised as outlined above (Fig.\ \ref{CLAs-ua-projection-figure}).
We call the first lens encountered by the light ray $L_1$, the second lens $L_2$.
Their centre positions are $\bm{C}_1$ and $\bm{C}_2$.
We define the unit vectors  $\hat{\bm{u}}$ and $\hat{\bm{v}}$ in the two cylinder-axis directions of the cylindrical lenses, and a unit vector $\hat{\bm{a}}$ that is perpendicular to both $\hat{\bm{u}}$ and $\hat{\bm{v}}$ and which points from the first lens to the second.
We use $\hat{\bm{u}}$, $\hat{\bm{v}}$ and $\hat{\bm{a}}$ as basis vectors for a an (not necessarily right-handed) orthogonal coordinate system.

We study light-ray transmission through one telescope generalised as outlined above in two orthographic projections, namely in the $(u, a)$ plane and in the $(v, a)$ plane.
Fig.\ \ref{CLAs-ua-projection-figure} shows the orthographic projection into the $(u, a)$ plane.
In this projection, the curvature of the lenses is such that the first lens, $L_1$, has focal length $f_{1,u}$ and the second lens, $L_2$, has focal length $f_{2,u}$.
The line $F_u$, which is a distance $f_{1,u}$ in the $a$ direction behind $L_1$, is the common focal line in this projection.
Any light ray entering the telescope through $L_1$ with direction $\bm{d}$ will intersect $F_u$ at the point $\bm{P}$ such that
\begin{align}
P_u = C_{1,u} + \frac{d_u}{d_a} f_{1,u}, \quad
P_a = C_{1,a} + f_{1,u}.
\label{P-ua-projection-equations}
\end{align}
We assume that the light ray leaves the telescope through $L_2$, which is the case of standard refraction \cite{Courtial-2009}.
The light ray will leave with direction $\bm{d}^\prime$ such that
\begin{align}
d^\prime_u
&= \frac{C_{2,u} - P_u}{f_{2,u}}
= \frac{C_{2,u} - C_{1,u}}{f_{2,u}} - \frac{d_u f_{1,u}}{d_a f_{2,u}},
\label{dPrime-ua-projection-u-equation-1} \\
d^\prime_a
&= \frac{C_{2,a} - P_a}{f_{2,u}}
= \frac{C_{2,a} - C_{1,a} - f_{1,u}}{f_{2,u}},
\label{dPrime-ua-projection-a-equation-1}
\end{align}
where we have divided by $f_{2,u}$ so that the light-ray direction is always pointing pointing outwards.
But as
\begin{align}
C_{2,u} = C_{1,u} + o_u, \quad
C_{2,a} = C_{1,a} + f_{1,u} + f_{2,u},
\end{align}
Eqns (\ref{dPrime-ua-projection-u-equation-1}) and (\ref{dPrime-ua-projection-a-equation-1}) become
\begin{align}
d^\prime_u
&= \frac{o_u}{f_{2,u}} - \frac{d_u f_{1,u}}{d_a f_{2,u}},
\label{dPrime-ua-projection-u-equation-2} \\
d^\prime_a &= 1.
\label{dPrime-ua-projection-a-equation}
\end{align}
In terms of the dimensionless quantities
\begin{align}
\delta_u = \frac{o_u}{f_{1,u}}, \quad
\eta_u = - \frac{f_{2,u}}{f_{1,u}}
\end{align}
we can write Eqn (\ref{dPrime-ua-projection-u-equation-2}) in the form
\begin{align}
d^\prime_u
= \frac{d_u / d_a - \delta_u}{\eta_u}.
\label{dPrime-ua-projection-u-equation}
\end{align}

A similar argument applied to the $(v, a)$ projection gives an equation for $d^\prime_v$ that is analogous to Eqn (\ref{dPrime-ua-projection-u-equation}), so that the full set of equations describing the generalised law of refraction for gCLAs is
\begin{align}
d^\prime_u = \frac{d_u / d_a - \delta_u}{\eta_u}, \quad
d^\prime_v = \frac{d_v / d_a - \delta_v}{\eta_v}, \quad
d^\prime_a = 1.
\label{dPrime-equations}
\end{align}
These equations can alternatively be written in the vector form
\begin{align}
\bm{d}^\prime
&= \frac{(\bm{d} \cdot \hat{\bm{u}}) / (\bm{d} \cdot \hat{\bm{a}}) - \delta_u}{\eta_u} \hat{\bm{u}}
+ \frac{(\bm{d} \cdot \hat{\bm{v}}) / (\bm{d} \cdot \hat{\bm{a}}) - \delta_v}{\eta_v} \hat{\bm{v}}
+ \hat{\bm{a}}.
\label{vector-form-generalised-law-of-refraction}
\end{align}
Eqns (\ref{dPrime-equations}) and (\ref{vector-form-generalised-law-of-refraction}) are the main results of this paper.


\begin{figure}
\begin{center}
\includegraphics{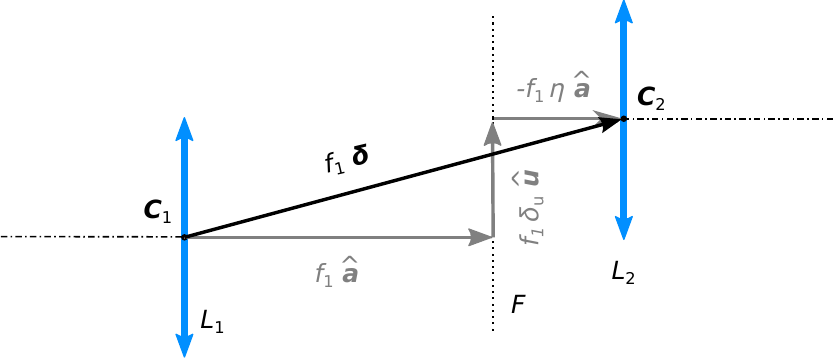}
\end{center}
\caption{\label{CLAs-eta-figure}Geometry of gCLAs for the case $\eta_u = \eta_v = \eta$, which occurs when $f_{1, u} = f_{1, v} = f_1$ and $f_{2, u} = f_{2, v} = f_2 = - \eta f_1$.
The drawing shows the orthographic projection into the $(u, a)$ plane.
There is now a single common focal plane, $F$, a distance $f_1$ behind $L_1$ and $f_2$ in front of $L_2$.
}
\end{figure}

We briefly discuss the special case $\eta_u = \eta_v = \eta$.
If we choose the length of the incident light-ray-direction vector, $\bm{d}$, such that $\bm{d} \cdot \hat{\bm{a}} = 1$, then Eqn (\ref{vector-form-generalised-law-of-refraction}) simplifies to
\begin{align}
\bm{d}^\prime
= \frac{\bm{d} - \bm{\delta}}{\eta},
\end{align}
where we have defined the vector
\begin{align}
\bm{\delta} = \left( \begin{array}{c} \delta_u \\ \delta_v \\ 1 - \eta \end{array} \right),
\end{align}
which can be related to the positions of the lens centres, $\bm{C}_1$ and $\bm{C}_2$, through the equation (see Fig.~\ref{CLAs-eta-figure})
\begin{align}
\bm{\delta} = \frac{\bm{C}_2 - \bm{C}_1}{f_1}.
\end{align}

\section{Implementation in open-source raytracer TIM}

\noindent
METATOYs are, at least in principle, very easy and cheap to mass-produce.
Confocal cylindrical lenslet arrays, for example, which refract light like Dove-prism arrays, can be essentially fabricated from the same lenticular arrays from which 3D postcards are made \cite{Blair-et-al-2009}.
However, sometimes building high-quality prototypes can be difficult and expensive.
To be able to ``experiment'' nevertheless with many different types of METATOYs, we have written our own custom raytracer called TIM (which stands for The Interactive METATOY) \cite{Lambert-et-al-2012,Oxburgh-et-al-2013}.

But TIM is more than a research tool.
TIM is written in Java and therefore easily distributed (it runs on any computer that supports the Java Virtual Machine (JVM) 1.6, and it can be run as a Java Application or as a Java Applet that can be embedded into web pages, although this is becoming increasingly difficult due to security restrictions).
TIM is also interactive, and arguably fun to play with (TIM can, for example, create anaglyphs and random-dot stereograms), and therefore ideally suited as a dissemination tool that can encourage exploration of aspects of our research.



\begin{figure}
\begin{tabular}{cc}
(a) & \includegraphics[width=0.9 \columnwidth]{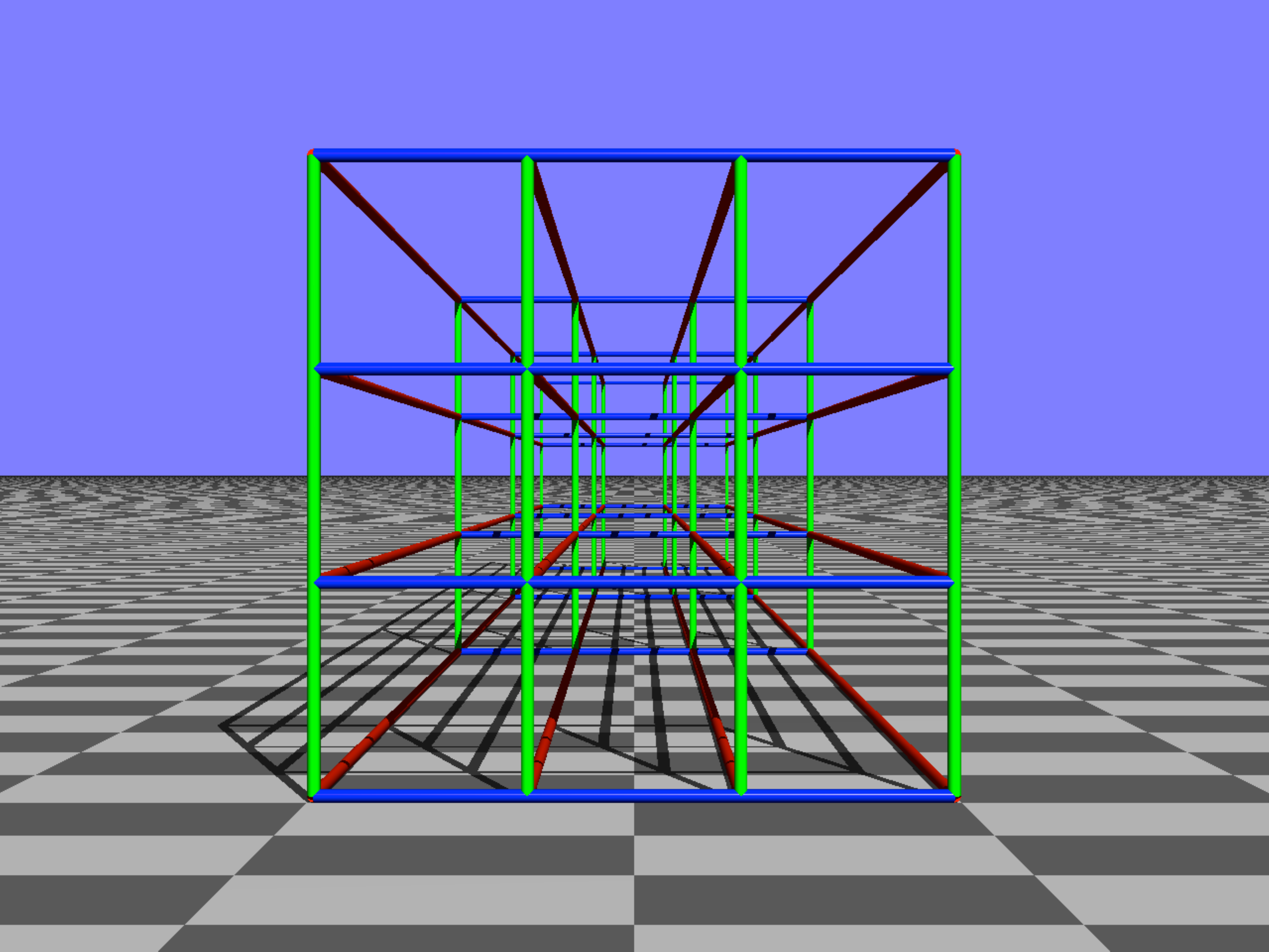} \\
(b) & \includegraphics[width=0.9 \columnwidth]{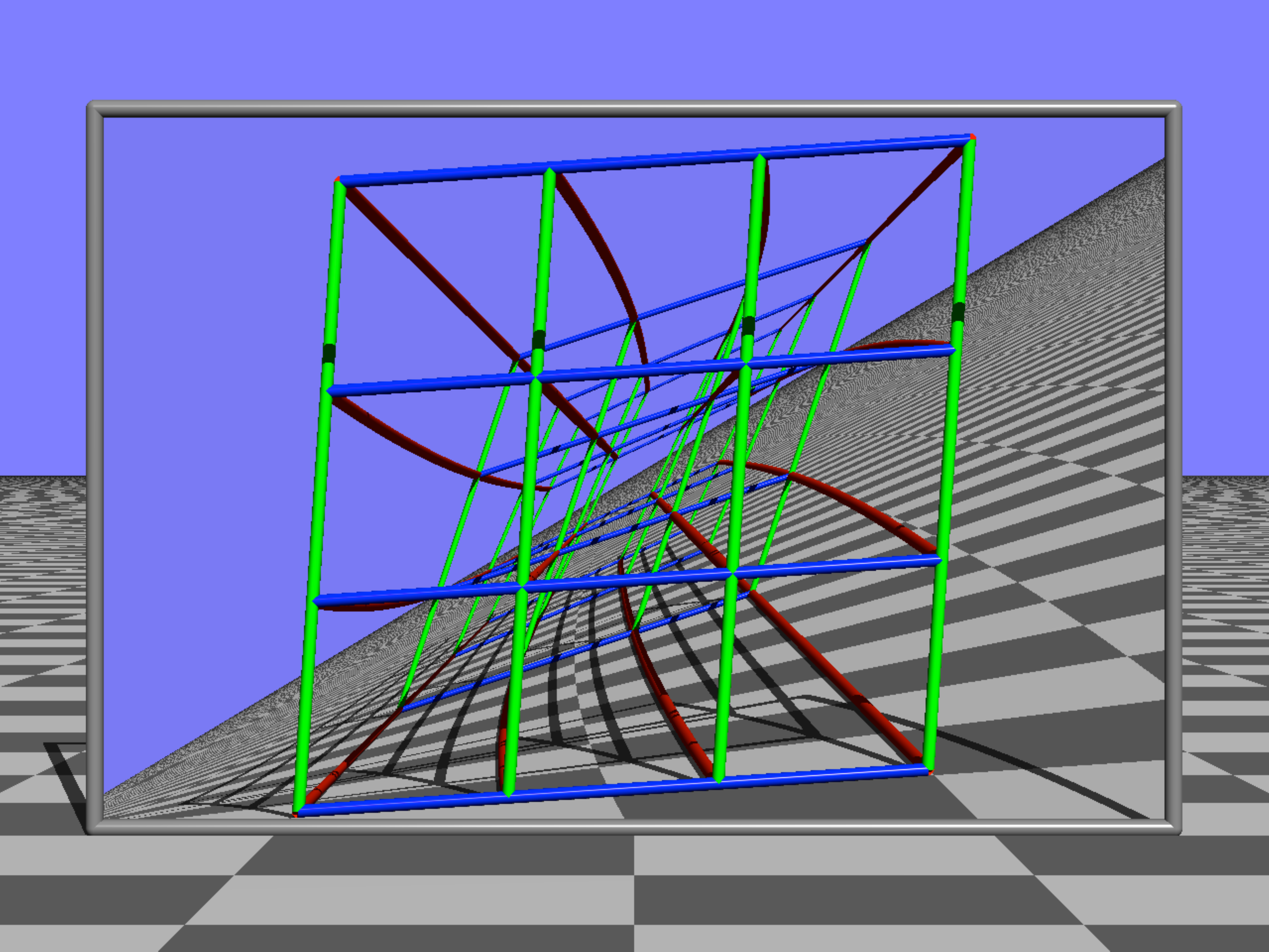} \\
(c) & \includegraphics[width=0.9 \columnwidth]{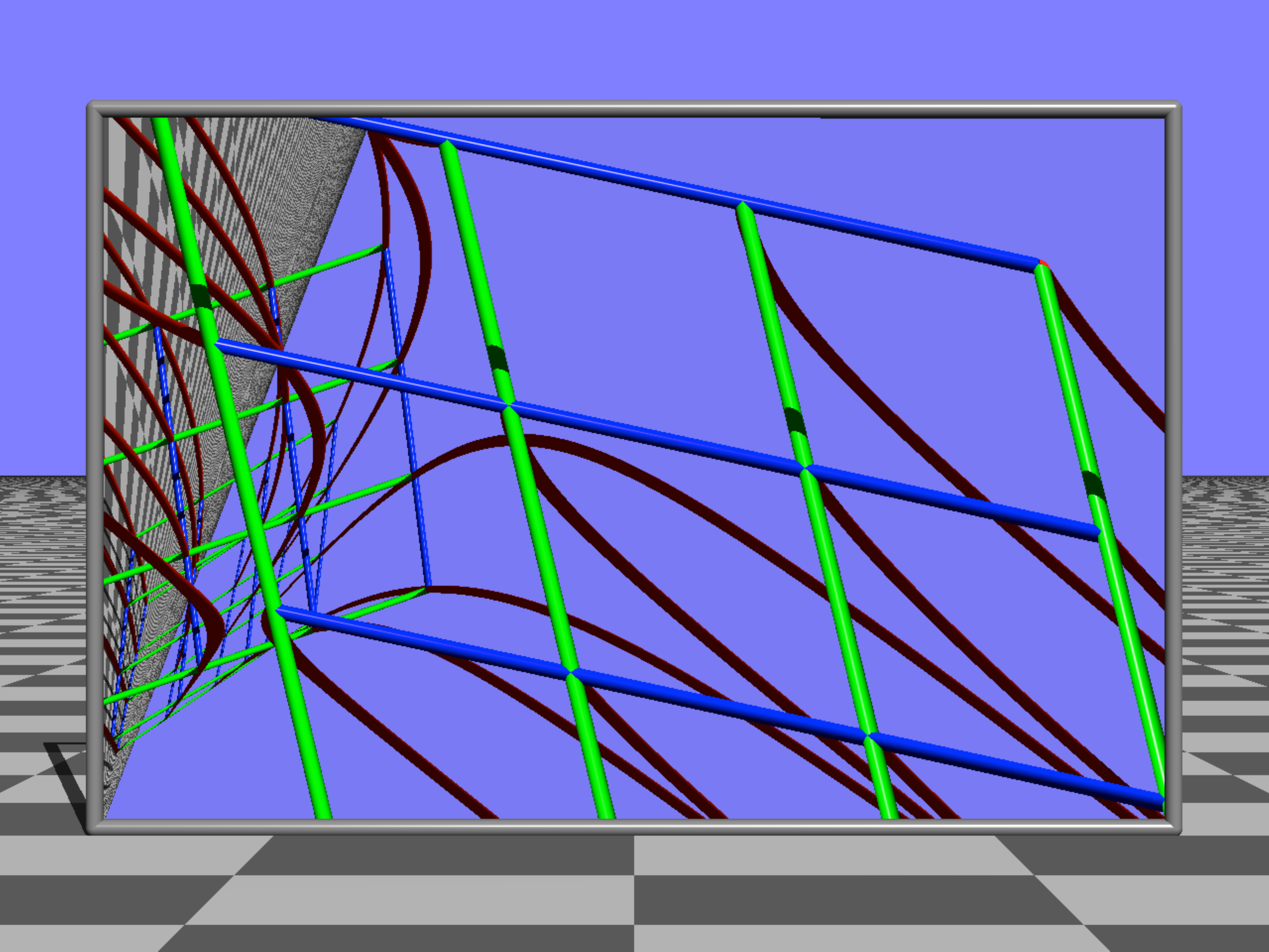}
\end{tabular}
\caption{\label{CLAs-view-figure}Simulated view of a lattice, on its own~(a) and seen through different generalised confocal lenslet arrays (b, c).
The input parameters were as follows:
(b)~$\bm{a} = (0, 0, 1)$, $\bm{u} = (1, 1, 0)$, $\eta_u = 4$, $\eta_v = 1$, $\delta_u = \delta_v = 0$; (c)~$\bm{a} = (-0.02, 0, 1)$, $\bm{u} = (0.79, 1, 0.02)$, $\eta_u = 2$, $\eta_v = -0.5$, $\delta_u = -0.2$, $\delta_v = 0.05$.
In both cases, the vector components are given in the basis of the object's local coordinate system.
The images were calculated using an extended version of the open-source raytracer TIM~\cite{Lambert-et-al-2012,Oxburgh-et-al-2013}.}
\end{figure}

We have now added to TIM the capability of ray tracing through gCLAs.
Fig.\ \ref{CLAs-view-figure} shows examples of simulated views through different gCLAs.
The gCLAs are described in terms of the numbers $\delta_u$, $\delta_v$, $\eta_u$, $\eta_v$, which are described above, and the vectors $\bm{a}$ and $\bm{u}$.
The vectors $\bm{a}$ and $\bm{u}$ do not need to be normalised; TIM does this internally.
It is not even necessary that the vector $\bm{u}$ as entered is perpendicular to $\bm{a}$, as TIM sets $\bm{u}$ to the part of the entered vector $\bm{u}$ that is perpendicular to $\bm{a}$.
The vector $\bm{v}$ is calculated as $\bm{v} = \bm{a} \times \bm{u}$, and is therefore automatically perpendicular to $\bm{a}$ and $\bm{u}$.

\section{Conclusions}

\noindent
We have derived here the law of refraction for gCLAs, which are very versatile components.
This law of refraction forms the basis of a number of potential applications, which we are currently investigating.
We have also extended our raytracer TIM to be able to simulate light propagation through gCLAs, allowing us to perform numerical experiments with these components.

The tools are now in place to investigate a number of potential applications of gCLAs.

\bibliographystyle{osajnl}
\bibliography{/Users/johannes/Documents/work/library/Johannes}

\end{document}